\documentclass[journal]{IEEEtran}

\usepackage{cite}
\usepackage{graphicx}
\usepackage{amsmath}
\usepackage{amssymb}
\usepackage{algorithm}
\usepackage{algorithmic}
\usepackage{color}

\begin{document}

\title{Coding vs. Spreading for Narrow-Band Interference Suppression}


\author{Xiaofu~Wu and Zhen~Yang 

\thanks{This work was presented in part at the ICC 2014, Sydney, Australia, 2014. This  work was supported in part by the National Science Foundation of China under Grants 61372123, 61271335, 61032004, by the Key University Science Research Project of Jiangsu Province under Grant 14KJA510003 and by the Scientific Research Foundation of NJUPT under Grant NY213002.}
\thanks{Copyright (c) 2015 IEEE. Personal use of this material is permitted. However, permission to use this material for any other purposes must be obtained from the IEEE by sending a request to pubs-permissions@ieee.org.}
\thanks{Xiaofu~Wu and Zhen~Yang are with the Key Lab of Ministry of Education in Broadband Wireless Communication and Sensor Network Technology, Nanjing University of Posts and Telecommunications, Nanjing 210003, China (Emails:xfuwu@ieee.org, yangz@njupt.edu.cn).}}


\maketitle

\begin{abstract}
The use of active narrow-band interference (NBI) suppression in direct-sequence spread-spectrum (DS-SS) communications has been extensively studied. In this paper, we address the problem of optimum coding-spreading tradeoff for NBI suppression.  With maximum likelihood decoding, we first derive upper bounds on the error probability of coded systems in the presence of a special class of NBI, namely, multi-tone interference with orthogonal signatures. By employing the well-developed bounding techniques, we show there is no advantage in spreading, and hence a low-rate full coding approach is always preferred. Then, we propose a practical low-rate turbo-Hadamard coding approach, in which the NBI suppression is naturally achieved through iterative decoding. The proposed turbo-Hadamard coding approach employs a kind of coded spread-spectrum signalling with time-varying spreading sequences, which is sharply compared with the code-aided DS-SS approach. With a spreading sequence of length 32 and a fixed bandwidth allocated for both approaches, it is shown through extensive simulations that the proposed turbo-Hadamard coding approach outperforms the code-aided DS-SS approach for three types of NBI, even when its transmission information rate is about 5 times higher than that of the code-aided DS-SS approach. The use of the proposed turbo-Hadmard coding approach in multipath fading channels is also discussed.
\end{abstract}


\IEEEpeerreviewmaketitle

\section{Introduction}
\PARstart{S}{pread-spectrum} signalling has found wide applications in both civilian and military communications. Although spread-spectrum communication is inherently resistant to narrow-band interference (NBI), which may be caused in a hostile environment or by coexistence with low-rate communications, substantial performance gains can be achieved through the use of active NBI suppression prior to despreading and demodulating \cite{Poor_NBI,Poor_codeI}. The NBI signal is often modeled either as a sinusoidal signal (tone) or an autoregressive (AR) signal. The NBI signal can also be a digital signal, such as would arise, for example, in a multirate signalling system\cite{Poor_NBI}.

In the past, a significant amount of research has been concerned with the development of techniques for active NBI suppression in spread-spectrum systems. In general, there are two categories for these techniques, one is based on linear signal processing techniques \cite{Milstein} and the other employs model-based techniques \cite{Poor_NBI}, including nonlinear filtering techniques and multiuser detection techniques. In \cite{Poor_codeI,Poor_codeII}, a code-aided approach was proposed by Poor and Wang for NBI suppression in direct-sequence code-division multiple-access (DS/CDMA) networks. This method, which is based on the minimum mean square error (MMSE) linear detector, has been shown to outperform all of the previous linear or nonlinear methods for NBI suppression. In recent years, NBI suppression has also been considered in ultra-wide bandwidth (UWB) communications \cite{Yuan,SongNBI,GomaaSparsity,ShaoPNDesign,AndreaNBI}. New signal processing algorithms for NBI suppression have also been reported in \cite{PerezNBI,LamareNBI}.

The code-aided direct-sequence spread-spectrum (DS-SS) approach \cite{Poor_codeI} can be combined with channel coding to further improve the system performance. However, this traditional use of channel coding  may not exploit its full potential for NBI suppression. For a spread-spectrum multiple-access channel, it was pointed out early by Viterbi \cite{ViterbiJSAC} that the ultimate capacity can be achieved when the entire spreading is dedicated to error control. Optimum coding-spreading tradeoff for a multiple-access channel was also addressed in \cite{VerduCodeSpread}. It was implied in \cite{VerduCodeSpread} that coding and spreading should be combined using low-rate codes with the maximized coding gain. Such a strategy has been considered in \cite{FrengerCodeSpread,LiPingMAC}. However, it remains open for the problem of optimum coding-spreading tradeoff when the NBI suppression is concerned.

The low-rate turbo-Hadamard codes, originally proposed in \cite{LiPingTH}, have been shown to perform very close to the ultimate Shannon limit for the additive white Gaussian noise (AWGN) channel. With a very simple trellis representation of component convolutional Hadamard codes, an iterative decoder was proposed in \cite{LiPingTH}, which can be efficiently implemented with negligible trellis complexity but leaving the complexity with the $a$ $posteriori$ probability (APP) decoding of the Hadamard code. It was also shown \cite{LiPingTH} that the APP decoding of Hadamard codes can be accomplished by a very efficient APP Fast-Hadamard-Transform (FHT). In \cite{LiPingMAC}, a low-rate turbo-Hadamard coding scheme was proposed to replace the traditional coding-spreading structure of DS-SS systems, which leads to significant performance improvement for the $synchronous$ multiple-access channel.

This paper focuses on the problem of coding versus spreading for the purpose of NBI suppression, a problem of great concern in military communications. For military communications, the signal must be designed so that it cannot be received and decoded by enemy forces, or it has low probability of interception (LPI). This is often achieved by spreading the signal so that its bandwidth is much larger than the true date rate. Hence, a low-rate coding structure for military communications is desired due to the requirement of LPI. For hostile jamming from potential enemy forces, the jammer often monitors the transmitted signal and concentrates its power in a few frequencies within the dominant transmitted signal bandwidth. Hence, the jamming signal is often regarded as NBI.

For spread-spectrum signalling aiming at NBI suppression, we conjecture that the entire spreading should be again dedicated to error control coding. In this paper, we first provide a partial theoretical justification of this conjecture under maximum-likelihood decoding. Then, we propose a practical low-rate turbo-Hadamard coding approach with consequently large bandwidth expansion for NBI suppression. We point out that this conjecture has also been independently observed in \cite{Elezabi} for partial-band jamming in frequency-hopping systems.

For the purpose of NBI suppression, we observe that if a full turbo-Hadamard codeword (without puncturing information bits for each component convolutional-Hadamard code) is transmitted over the channel, the resulting signal presents the form of DS-SS signalling, but with time-varying Hadamard sequences. Therefore, it can be well understood that the code-aided approach for NBI suppression might be still useful. Due to  a low-rate coding structure, the NBI suppression can be naturally achieved through iterative decoding.

In \cite{SchlegelCL,CaiTCOM}, it was observed that a proper use of interleaving in direct-sequence spreading can be beneficial for multi-user interference (MUI) cancellation, where  the spreading sequence is time-varying due to the introduction of interleaving.  The proposed turbo-Hadamard coding approach employs a kind of DS-SS signalling with time-varying spreading sequences. Compared to \cite{SchlegelCL,CaiTCOM}, the observed time-varying spreading sequences in the turbo-Hadamard coding approach are controlled by the coding rule, and hence an extra coding gain can be achieved.

The main contributions of this paper are summarized as follows:
\begin{enumerate}
    \item
    By assuming a special type of NBI, namely, the multi-tone interference with orthogonal signatures, we derive upper bounds on the error probability of coded systems in the presence of both AWGN and NBI. Then, we provide a partial justification of the conjecture that the optimum coding-spreading tradeoff always favors full coding for NBI suppression.
    \item
    A practical low-rate full coding approach for NBI suppression is proposed, which has never been reported for the purpose of NBI suppression.
    \item
    With a matched filter receiver at chip rate, the proposed turbo-Hadamard coding approach shows significantly higher spectral efficiency compared to the traditional DS-SS approach.
    \item
    Considering that in the presence of NBI, the extrinsic information for iterative decoding cannot be directly extracted at the positions of systematic (information) bits, an approximate method is proposed and simulations show that this approximate method works well.
    \item
    It is shown that the APP decoding of Hadamard codes in the presence of NBI can be implemented using FHT, which facilitate its practical implementation.
    \item
    The full coding approach for NBI suppression can also be applied to multipath fading channels. We consider quasi-static fading channels and show that it is possible to develop a low-complexity iterative receiver for multipath channels with long Hadamard sequences.
\end{enumerate}

The rest of the paper is organized as follows. The optimum coding-spreading tradeoff for NBI suppression is investigated in Section II.  In Section-III, we give a brief description of low-rate full turbo-Hadamard codes without puncturing. This coding approach is employed for NBI suppression through iterative decoding as shown in Section-IV. Section-V presents simulation results and Section-VI concludes the paper.

\begin{figure*} 
   \centering
   \includegraphics[width=0.82\textwidth]{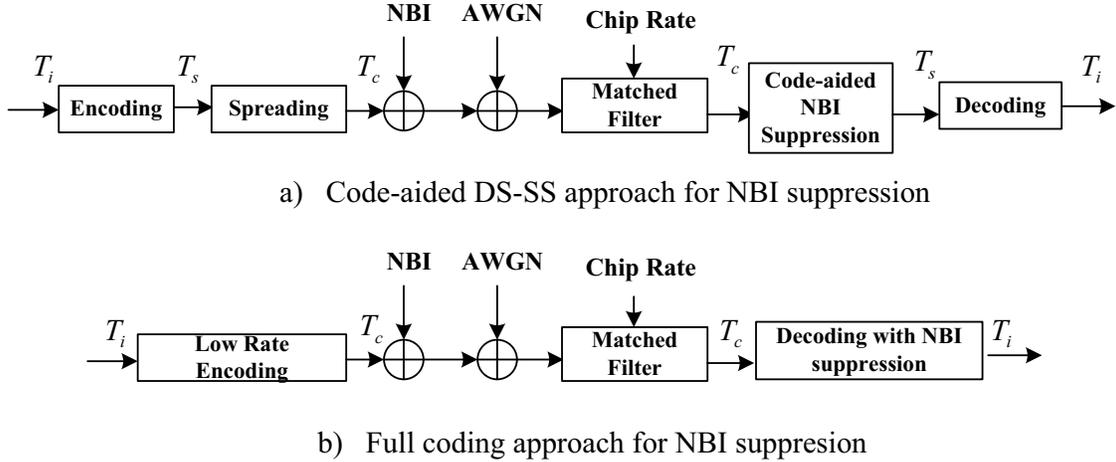}
   \caption{The block diagrams of two NBI suppression approaches.}
   \label{fig:codingView}
\end{figure*}
\section{Coding vs. Spreading}
\subsection{System Model}
Consider a baseband signal model with only one user at the receiver
\begin{equation}
   r(t) = \sqrt{P_s} c(t) + i(t) + w(t),
\end{equation}
where $c(t)$ is the spread-spectrum signal of the user and $P_s$ is the received signal power, $i(t)$ and $w(t)$ represent the NBI signal and the ambient channel noise, respectively.

We consider two approaches for NBI suppression, namely, a code-aided DS-SS approach and a full coding approach, as shown in Fig. \ref{fig:codingView}. For both approaches, it is assumed that a block channel coding scheme is employed. At the transmitter, information bits are grouped into data frames and then input to the block encoder.  For a full coding approach, the coding rate should be low enough for expanding the signal spectrum. For a coded DS-SS approach, the (channel) coding rate is often larger than $\frac{1}{2}$ and the signal spectrum is mainly expanded by direct-sequence spreading.

It should be mentioned that the conventional coded DS-SS approach can also be viewed as a general coding approach, where the traditional channel coding is serially concatenated with repetition coding (direct-sequence spreading) as described in Fig. \ref{fig:codingView}.  Hence, both approaches can be viewed as a unified coding approach.

Denote by $T_F$ and $T_i$, the frame duration and the information bit duration, respectively. Suppose now that every $K$ information bits are grouped as a date frame, namely, $T_F = K T_i$. With a unified coding view for both approaches shown in Fig. \ref{fig:codingView}, we can consider that information bits are input to a low-rate encoder, which outputs coded bits (in chips) with duration $T_c$ (chip duration). In this paper, we always assume that $T_i \gg T_c$.

Assuming that a block channel coding scheme is independently implemented over each frame, we can arrive at a signal vector representation, by projection of the signals onto a finite basis. Indeed, the received signal (column) vector $\mathbf{r}$ of length $L=T_F/T_c$ during a frame can be obtained by passing $r(t), t\in [0,T_F)$ through a chip-matched filter, followed by a chip-rate sampler, giving
\begin{eqnarray}
    \mathbf{r} =\sqrt{P_s}\mathbf{c} + \mathbf{i} + \mathbf{w},
\end{eqnarray}
where $\mathbf{c}$ is a binary-phase-shift-keying (BPSK) modulated codeword, $\mathbf{w}$ is a sample vector of zero-mean Gaussian noise with variance $\sigma^2$ and $\mathbf{i}$ denotes a collected column vector form of the NBI.

For the traditional coded DS-SS approach, this model still holds. However, the NBI suppression is often achieved via linear MMSE approach with one-shot symbol duration $T_s$, where $n=T_s/T_c$ is often employed to denote the spreading factor. Hence, the received signal vector should be further partitioned as sub-vectors of length $n$, i.e., $\mathbf{r}=[\mathbf{r}_1^T,\mathbf{r}_2^T,\cdots,\mathbf{r}_k^T,\cdots,\mathbf{r}_B^T]^T$ with $B=T_F/T_s$. Then, the $k$th received sub-vector during one symbol period $T_s$ can be represented as
\begin{eqnarray}
   \label{vec_ch}
    \mathbf{r}_k =\sqrt{P_s}\mathbf{c}_k + \mathbf{i}_k + \mathbf{w}_k,
\end{eqnarray}
where $\mathbf{c}_k$ is the data-bearing ($\pm 1$ modulated) spreading-sequence vector, $\mathbf{w}_k$ is a Gaussian noise vector and $\mathbf{i}_k$ denotes a NBI vector, which is assumed to be wide-sense stationary with mean zero and covariance matrix $R_i$.

With a linear MMSE approach for NBI suppression, the detector simply makes a symbol-by-symbol decision as $\hat{b}_k=\text{sgn}\left(\mathbf{u}^T\mathbf{r}_k\right)$, where $\mathbf{u}$ is chosen to minimize the mean-square error
\begin{eqnarray}
  \label{eq:MMSE}
    MSE \triangleq E\left\{\mathbf{u}^T\mathbf{r}_k - \text{sgn}\left(\mathbf{u}^T \mathbf{r}_k\right)\right\}.
\end{eqnarray}

\subsection{Maximum-Likelihood Decoding under Gaussian-distributed NBI}
As shown in Fig. \ref{fig:codingView}, both approaches can be considered as a block coding scheme, where a linear binary $(L,K)$ block code $C$ is employed in the sense of chip-rate transmission with the coding rate given by
\begin{equation}
    R_c=\frac{K}{L}=\frac{T_c}{T_i}.
\end{equation}

Hence, one can bound the error probability of this unified coding system in the presence of NBI under maximum-likelihood (ML) decoding to show the fundamental tradeoff of coding versus spreading.

For ML decoding, one has to determine the codeword $\mathbf{c}$ for which the conditional probability density function
$p\left(\mathbf{r}|\mathbf{c}\right)$ reaches its maximum.

In general, NBI is a stochastic process with infinite memory. In practice, it can be well approximated by finite memory. For ease of processing, a block-independent approximation is often employed, where the NBI process $\mathbf{i}$ can be partitioned into a sub-block vector form of $\mathbf{i}=[\mathbf{i}^T_1,\mathbf{i}^T_2,\cdots,\mathbf{i}^T_B]^T$ with each sub-vector $\mathbf{i}_k$ of length $n=T_s/T_c$ \footnotemark\footnotetext{It is assumed that the size of memory coincides well with the symbol period in the DS-SS approach for ease of processing.} satisfying $E\{\mathbf{i}_k \cdot \mathbf{i}^T_l\}=\delta_{k,l} R_i$.  Note that there is still some room for performance enhancement if one considers a longer memory size ($>n$) for NBI, which clearly compromises with increased decoding complexity.

Then, $p\left(\mathbf{r}|\mathbf{c}\right)$ can be factored as
\begin{eqnarray}
\label{eq:pr0}
    p\left(\mathbf{r}|\mathbf{c}\right) = \prod_{k=1}^B p\left(\mathbf{r}_k|\mathbf{c}_k\right).
\end{eqnarray}

To evaluate $p\left(\mathbf{r}_k|\mathbf{c}_k\right)$, it requires to know the probability distribution of $\mathbf{i}_k$. Here, we assume that $\mathbf{i}_k$ is a Gaussianly-distributed vector, namely, $\mathbf{i}_k \sim  \mathcal{N}(\mathbf{0}, R_i)$. If the NBI is generated by an AR process, this is just a precise assumption. If the NBI is modeled as a sinusoidal signal with multiple tones, or the digital NBI signal, this is approximately satisfied.
With this assumption, (\ref{eq:pr0}) can be clearly represented as
\begin{eqnarray}
\label{eq:pr}
    p\left(\mathbf{r}|\mathbf{c}\right) \propto \exp\left(-\frac{1}{2} \sum_{k=1}^B\left(\mathbf{r}_k - \mathbf{c}_k\right)^T Q^{-1}\left(\mathbf{r}_k - \mathbf{c}_k\right) \right),
\end{eqnarray}
where $P_s$ is assumed to 1 and $Q = R_i + \sigma^2 I$.

Hence, an ML decoder only needs to choose a coded
sequence $\mathbf{c}$ which maximizes the metric
\begin{eqnarray}
\label{eq:eta}
   \eta = \sum_{k=1}^B \left<\mathbf{c}_k, \mathbf{y}_k\right> +\lambda(\mathbf{c}).
\end{eqnarray}
where $\mathbf{y}_k= Q^{-1}\mathbf{r}_k$ and
\begin{equation}
    \lambda(\mathbf{c}) = -\sum_{k=1}^B\frac{1}{2} \mathbf{c}^T_k Q^{-1} \mathbf{c}_k.
\end{equation}

\subsection {Error Probability under Multi-tone Interference with Orthogonal Signatures}
The pair-wise error probability (PEP),  which represents the probability of choosing the coded
sequence $\hat{\mathbf{c}}$ when the coded sequence $\mathbf{c}$
was transmitted, is the fundamental expression for the
construction of the union or other upper bounds on the average
error probability performance of a system.

In general, it is difficult to derive the PEP in the presence of NBI. To gain essential insights, we, however, focus on a special class of NBI process, namely, multi-tone interference with orthogonal signatures \cite{Poor_codeI}. This class of NBI process is approximately Gaussian-distributed when the number of tones is large thanks to the central limit theorem. Hence, the metric (\ref{eq:eta}) still holds with ML decoding.

Letting $\eta$ (or $\hat{\eta}$) denote the metric
(\ref{eq:eta}) for the correct (or incorrect) coded sequence $\mathbf{c}$ (or
$\hat{\mathbf{c}}$), then the PEP of choosing the coded sequence $\hat{\mathbf{c}}$ instead of the actual transmitted
coded sequence $\mathbf{c}$ is given by
\begin{equation}
\label{eq:8}
  P\left(\mathbf{c}\rightarrow \hat{\mathbf{c}} \right) = \text{Pr}\left( \eta - \hat{\eta} < 0 \bigg{|}\mathbf{c} \right).
\end{equation}

The metric difference can be computed as
\begin{equation}
    \eta - \hat{\eta} = \sum_{k=1}^B \left<\mathbf{c}_k - \hat{\mathbf{c}}_k, \mathbf{y}_k\right> + \lambda(\mathbf{c}) - \lambda(\hat{\mathbf{c}}),
\end{equation}
where
\begin{equation}
   \lambda(\mathbf{c}) - \lambda(\hat{\mathbf{c}}) =  - \frac{1}{2} \sum_{k=1}^B \left(\mathbf{c}_k - \hat{\mathbf{c}}_k \right)^T Q^{-1} \left(\mathbf{c}_k + \hat{\mathbf{c}}_k\right).
\end{equation}

For analytic tractability, we should further introduce a series of approximations as did in the Appendix. For conciseness, the derivation is summarized in the Appendix.

Let $h=d_H\left(\hat{\mathbf{c}},\mathbf{c}\right)$ be the Hamming distance between $\hat{\mathbf{c}}$ and $\mathbf{c}$, and $\gamma_l, l=1,\cdots,F$, be the $l$th tone's power-to-noise ratio. In the Appendix, it is shown that
\begin{eqnarray}
    \mu_z = E\left\{\hat{\eta} - \eta\Big| \mathbf{c} \right\} \approx 2\sigma^{-2}\left(1-\sum_{l=0}^F \frac{\gamma_l}{1+n \gamma_l}\right) h ,
\end{eqnarray}
and
\begin{eqnarray}
    \sigma_z^2 = D\left\{\hat{\eta} - \eta\Big| \mathbf{c} \right\} \approx 4\sigma^{-2} \left(1-\sum_{l=0}^F \frac{\gamma_l}{1+n \gamma_l}\right) h
\end{eqnarray}
for sufficiently large $B$.

We now include the effect of $P_s$. Defining $\gamma_s = \frac{P_s}{2\sigma^2}$, we have
\begin{equation}
\label{eq:pep}
  P\left(\mathbf{c}\rightarrow \hat{\mathbf{c}} \right) \approx  Q\left(\sqrt{2\gamma_s \left(1-\sum_{l=0}^F \frac{\gamma_l}{1+n \gamma_l}\right) h}\right),
\end{equation}
where $Q(\cdot)$ is the Gaussian integral function
\begin{equation*}
  Q(x)=\frac{1}{\sqrt{2\pi}} \int_x^{+\infty} \exp\left(-\frac{t^2}{2}\right)dt.
\end{equation*}

With simple union bounding technique\cite{Divslar}, the word error probability can be bounded as
\begin{equation}
\label{eq:fer}
  P_w \le \sum_{h\ge h_{\min}} A_h Q\left(\sqrt{2\gamma_s \left(1-\sum_{l=0}^F \frac{\gamma_l}{1+n \gamma_l}\right) h}\right),
\end{equation}
where $A_h$ denotes the number of codewords of weight $h$ in $C$. Note that the union bound on bit error probability can also be obtained by employing the input-output weight enumerator \cite{Divslar}.

\subsection{Coding vs. Spreading}
\begin{figure}[htbp]
   \includegraphics[width=0.5\textwidth]{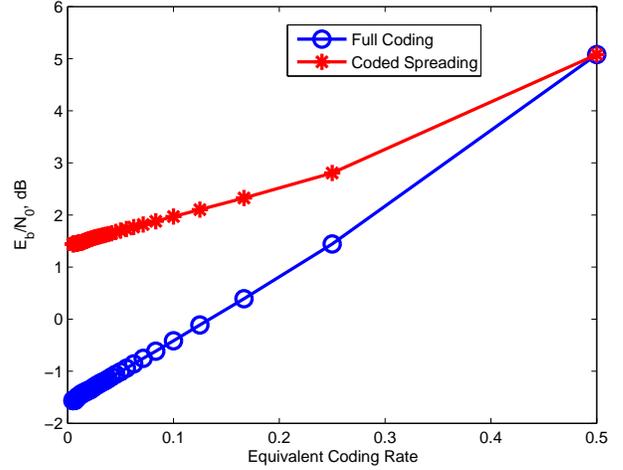} 
   \caption{Comparison of minimum threshold of full coding vs. coded spreading for random codes as $L\rightarrow \infty$.}
   \label{fig:thr}
\end{figure}

In this subsection, we still focus on the multi-tone interference with orthogonal signatures, as it admits a closed-form expression for the upper bound on the pair-wise error probability (\ref{eq:pep}).

With BPSK modulation, it is clear that $\gamma_s=R_c E_b/N_0$, where  $E_b$ is the information bit energy and $N_0/2$ is the two-sided power spectral density of the white Gaussian noise process at the receiver.

Define $\kappa_i=1-\sum_{l=0}^F \frac{\gamma_l}{1+n \gamma_l}$, $\delta = h/L, \mu(\delta)=(\ln A_h )/L$ and
\begin{equation}
  c_0(\text{Divsalar}) = \max_\delta \left\{\left(1-e^{-2\mu(\delta)}\right)\frac{1-\delta}{2\delta}\right\}.
\end{equation}
In \cite{Divslar}, Divsalar derived a simple upper bound and further showed that, if $E_b/N_0 >  \frac{1}{\kappa_i R_c} c_0(\text{Divsalar})$, the simple upper bound goes to 0 as the block size $L$ goes to infinity.

For the set of random linear $(L,K)$ block codes of rate $R_c$, it is well known that
\begin{equation}
   \mu(\delta) = H(\delta) - (1-R_c) \ln 2,
\end{equation}
where $H(\delta)=-\delta \ln \delta -(1-\delta) \ln(1-\delta)$.

Now, we are in a position to discuss the optimum coding-spreading tradeoff for NBI suppression. With a unified coding view, a coded DS-SS approach can be viewed as a serially-concatenated coding approach, where direct-sequence (of length $n$) spreading can be regarded as a simple $(n, 1)$ inner repetition coding scheme.

It is well known that the serial concatenation of any outer code with an inner repetition code can not produce any good code (namely, the weight distribution $A_h, h\ge 0$ is not desirable). For example, let us consider a powerful rate-1/2 outer code with $\mu_o(\delta)=H(\delta)-\frac{1}{2} \ln 2$, which is concatenated with an $(n, 1)$ inner repetition code. Then, the concatenated code has $\mu(\delta)=\frac{1}{n}\mu_o(n\delta)$. By employing the simple bound \cite{Divslar}, the minimum $E_b/N_0$ threshold conditioned on a given value of $\kappa_i$ can be computed as ($L \rightarrow \infty$)
 \begin{equation}
   \left(\frac{E_b}{N_0}\right)_{\text{threshold}} = \frac{1}{\kappa_i R_c} c_0(\text{Divsalar}) .
\end{equation}

In the case of $F=1, \gamma_l = 20$ dB, the minimum $E_b/N_0$ threshold versus the coding rate is plotted in Fig. \ref{fig:thr} for both NBI suppression approaches. For the coded spreading approach, we have employed a rate-1/2 random code as the outer code, which is further concatenated with a $(n,1)$ repetition code (spreading). Therefore, an equivalent coding rate of $\frac{1}{2n}$ is obtained. For the full coding approach,  the rate-$\frac{1}{2n}$ random code is employed.  As shown in Fig. \ref{fig:thr}, the optimum coding-spreading tradeoff always favors negligible spreading when $L\rightarrow \infty$.

Therefore, we arrive at a conclusion that the entire bandwidth expansion should be dedicated to error control for NBI suppression and a low-rate full coding approach designed with the maximized coding gain could be better than the code-aided DS-SS approach \footnotemark\footnotetext{In essence, this section provides just a partial justification of this claim due to a series of approximations and the restriction on multi-tone interference. Hence, a rigourous proof is still missing.}. In the remainder of this paper, we show that this conclusion is also supported in practice through extensive simulations for three types of NBI.

\section{Low-Rate Full Turbo-Hadamard Codes}
\begin{figure*} 
   \centering
   \includegraphics[width=0.72\textwidth]{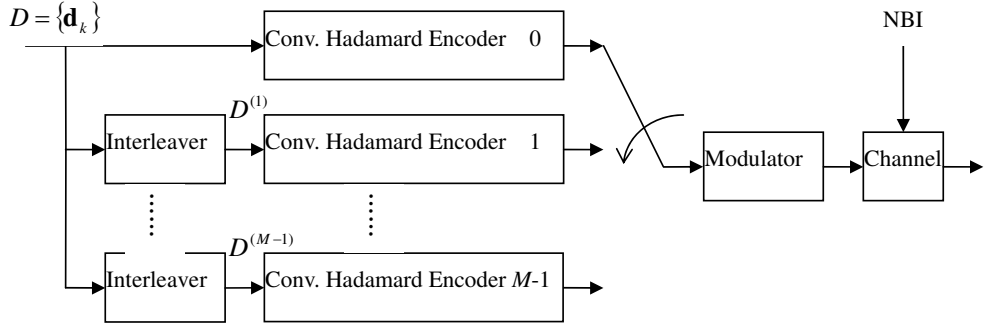}
   \caption{System Model.}
   \label{fig:sys}
\end{figure*}

For the purpose of NBI suppression, we employ a low-rate $full$ turbo-Hadamard code for spreading the signal bandwidth. When such a codeword is transmitted over a channel with BPSK signalling, it can be viewed as the employment of coded spread-spectrum signalling with time-varying spreading sequences, which competes well with the traditional DS-SS signalling for NBI suppression. It should be mentioned that the original turbo-Hadmard codes proposed in \cite{LiPingTH} use punctured Hadamard sequences for each codeword, which gives an obstacle for developing iterative decoding in the presence of NBI. Without puncturing, it also benefits a potential signal acquisition mechanism for the proposed full coding approach (with time-varying spreading sequences).

\subsection{Hadamard Codes}
An $n\times n$ Hadamard matrix with $n=2^r$ can be recursively constructed as
\begin{eqnarray}
 \label{eq:5}
H_n= \left[\begin{array}{cc} +H_{n/2} & +H_{n/2} \\
 +H_{n/2} & -H_{n/2}
 \end{array}\right],
\end{eqnarray}
and $H_1=[+1]$.

A length-$2^r$ bi-orthogonal Hadamard code carries $r+1$ information bits. Therefore, the rate of a length-$2^r$ Hadamard code is $R_h=\frac{r+1}{2^r}$.
For such a code, the codeword set is formed by the columns of $\pm H_n$, denoted by $\{\pm \mathbf{h}^j, j=0,1,\cdots,2^r-1\}$. In
a systematic encoding, the bit indexes $\mathcal{I}=\{0,1,2,4,\cdots,2^{r-1}\}$ are used as information bit positions.

\subsection{Convolutional-Hadamard Codes}
To encode a convolutional-Hadamard code, the information bit stream is first segmented into blocks of $r$ bits. The $k$th block can be written as
\begin{equation}
  \mathbf{d}_k=\left[d_{kr},d_{kr+1},\cdots,d_{kr+r-1}\right]^T,
\end{equation}
where $d_{kr+j}\in \{0,1\}, j=0,1,\cdots, r-1$. Then, each block is employed to produce a parity-check bit. For the $k$th block, the produced parity-check bit is
\begin{equation}
    p_k=d_{kr}\oplus d_{kr+1} \oplus \cdots \oplus d_{kr+r-1}.
\end{equation}
The parity bits of information blocks $\{p_k\}$ are further input to a rate-1 recursive convolutional code to produce the coded bits $\{q_k\}$. The simplest rate-1 recursive convolutional code is a two-state convolutional code with generator $1/(1+D)$, namely, the binary accumulator or the differential encoder, where the $k$th output coded bit is
\begin{equation}
  q_k = q_{k-1} \oplus p_k.
\end{equation}

The information blocks $\{\mathbf{d}_k\}$ are then appended with $\{q_k\}$ to form the augmented input blocks $[\mathbf{d}_k^T, q_k]^T$. The augmented blocks, each with $r+1$ bits, are input to the encoder of a Hadamard code, producing $n=2^r$ coded bits for each block.

\subsection{Full Turbo-Hadamard Codes}
A full turbo-Hadamard code is constructed by concatenating several convolutional-Hadamard codes in parallel.
Let $D=[\mathbf{d}_0,\mathbf{d}_1,\cdots,\mathbf{d}_{L-1}]$ be a group of information bits, which are input to the encoder for producing a full turbo-Hadamard codeword.
As shown in Fig. \ref{fig:sys}, $D$ ($=D^{(0)}$) and its interleaved versions $D^{(m)}, m=1,\cdots,M-1$ are employed by the component convolutional Hadamard codes to produce $M$ Hadamard codewords $C^{(m)}=[\mathbf{c}_{mL},\mathbf{c}_{mL+1},\cdots,\mathbf{c}_{mL+L-1}], m=0,\cdots,M-1$ of size $2^r\times L$. Hence, a full turbo-Hadamard codeword can be obtained by collecting all component codewords, namely, $C=[C^{(0)},C^{(1)},\cdots,C^{(M-1)}]=[\mathbf{c}_0,\mathbf{c}_1,\cdots,\mathbf{c}_{ML-1}]$, where $\mathbf{c}_k, k=0,1,\cdots,ML-1$, is the column vector of length $2^r$.

Compared to the original turbo-Hadamard codeword \cite{LiPingTH}, a full turbo-Hadamard codeword consists of $M$ sub-codewords, produced by $M$ component convolutional-Hadamard codes without puncturing. Hence, information bits can be repeatedly observed in the final codeword. The employment of full component Hadamard codewords can facilitate NBI suppression without sacrificing much bandwidth since the rate of each convolutional-Hadamard code is very low. With a full turbo-Hadamard code using BPSK modulation, the transmitted signal can be seen as a kind of direct-sequence spread-spectrum signalling with time-varying spreading-sequences (Hadamard sequences) of fixed length $n=2^r$.

The rate of a full turbo-Hadamard code is simply given by
\begin{equation}
   R_F = \frac{r}{M 2^r},
\end{equation}
while the original turbo-Hadamard code is of rate
\begin{equation}
   R_{TH} = \frac{r}{r+M(2^r-r)}.
\end{equation}
Hence, the ratio is
\begin{equation}
    \frac{R_F}{R_{TH}}= \frac{M 2^r- (M-1)r}{M 2^r},
\end{equation}
which quickly approaches 1 as $r$ increases.

\section{Iterative Decoding in the Presence of both NBI and AWGN}
We have shown in Section-II the potential of a full coding approach in NBI suppression with ML decoding. However, the complexity issue makes an ML decoder impractical. In this section, we develop a practical iterative decoding algorithm in the presence of both NBI and AWGN for a low-rate full turbo-Hadamard coding approach.

\subsection{Channel Model}
With a vector representation of the channel model (\ref{vec_ch}), we assume that the NBI is a block-independent random process.
For convenience, it is always assumed that the block memory length of NBI is the same as the length of component Hadamard codeword ($n=2^r$).

Therefore, when a full turbo-Hadamard codeword $C=[\mathbf{c}_0,\mathbf{c}_1,\cdots,\mathbf{c}_{ML-1}]$ is transmitted, the received equivalent complex base-band signal vector in the duration of Hadamard codeword $n T_c$ can be represented as (\ref{vec_ch}), where $\mathbf{c}_k$ is now the $k$th (component) Hadamard codeword of a full turbo-Hadamard codeword.

In what follows, three types of NBI signals \cite{Poor_codeI}, multi-tone signal, autoregressive (AR) signal, and digital signal, are considered.
For conciseness, we still use $P_s = 1$ in this section.

\subsection {APP Decoding of Hadamard Code in the Presence of NBI}
Assume now that all of the Hadamard codewords have independent and equal probabilities of occurrence. Let $\mathbf{c}_k=\left[c_k[0],c_k[1],\cdots,c_k[n-1]\right]^T$ be the transmitted
$k$th Hadamard codeword and $\mathbf{r}_k$ be the noisy observation of $\mathbf{c}_k$ (\ref{vec_ch}). The APP decoding of a Hadamard code over $\{+1,-1\}$ involves evaluating
the logarithm-likelihood ratio (LLR)
\begin{eqnarray}
\label{eq:llr}
    L_k[i] = \log \frac{P_r(c_k[i]=+1|\mathbf{r}_k)}{P_r(c_k[i]=-1|\mathbf{r}_k)} = \log \frac{\displaystyle \sum_{c_k[i]=+1} p\left(\mathbf{r}_k|\mathbf{c}_k\right)}{\displaystyle \sum_{c_k[i]=-1} p\left(\mathbf{r}_k|\mathbf{c}_k\right)}, \nonumber \\ i=0,1,\cdots, 2^r-1.
\end{eqnarray}

Let us define
\begin{eqnarray}
\label{eq:inner}
    \psi(\pm \mathbf{h}^j)&=&\exp\left(\pm \left<\mathbf{h}^j, \mathbf{y}_k\right>\right) \cdot \exp\left(-\frac{1}{2} {\mathbf{h}^j}^T Q^{-1} \mathbf{h}^j\right),\nonumber \\
                          &=&\exp\left(\pm \left<\mathbf{h}^j, \mathbf{y}_k\right> + \lambda_j\right),
\end{eqnarray}
where
\begin{equation}
\label{eq:lambda}
    \lambda_j = -\frac{1}{2} {\mathbf{h}^j}^T Q^{-1} \mathbf{h}^j.
\end{equation}

Then, (\ref{eq:llr}) can be expanded as shown in the top of the next page.
\newcounter{mytempeqncnt}
\begin{figure*}[!t]
\normalsize
\setcounter{mytempeqncnt}{\value{equation}}
\begin{eqnarray}
\label{eq:expand}
    L_k[i]  &=& \log \frac{\displaystyle\sum_{H[i,j]=+1} \exp\left(-\frac{1}{2} \left(\mathbf{r}_k - \mathbf{h}^j\right)^T Q^{-1}\left(\mathbf{r}_k - \mathbf{h}^j\right) \right) + \sum_{H[i,j]=-1} \exp\left(-\frac{1}{2} \left(\mathbf{r}_k + \mathbf{h}^j\right)^T Q^{-1}\left(\mathbf{r}_k + \mathbf{h}^j\right) \right)} {\displaystyle\sum_{H[i,j]=-1} \exp\left(-\frac{1}{2} \left(\mathbf{r}_k - \mathbf{h}^j\right)^T Q^{-1}\left(\mathbf{r}_k - \mathbf{h}^j\right) \right) + \sum_{H[i,j]=+1} \exp\left(-\frac{1}{2} \left(\mathbf{r}_k + \mathbf{h}^j\right)^T Q^{-1}\left(\mathbf{r}_k +\mathbf{h}^j\right) \right)}, \nonumber \\
    &=& \log \frac{\displaystyle\sum_{H[i,j]=\pm 1} \psi(\pm \mathbf{h}^j)}{\displaystyle\sum_{H[i,j]=\mp 1} \psi(\pm \mathbf{h}^j)}.
\end{eqnarray}
\setcounter{equation}{\value{mytempeqncnt}}
\hrulefill
\vspace*{4pt}
\end{figure*}
\addtocounter{equation}{1}

In the absence of NBI, it can be shown that $\lambda_j=-\frac{1}{2} n\frac{1}{\sigma^2}$ since $\sigma^2 {\mathbf{h}^j}^T Q^{-1}\mathbf{h}^j = \|\mathbf{h}^j \|^2=n$.  Efficient techniques for evaluating (\ref{eq:inner}) and (\ref{eq:expand}) were proposed in \cite{LiPingTH} using FHT.

In the presence of NBI, we now show that the FHT-based efficient techniques still work with the assumption that the NBI process is stationary.

For a stationary NBI process, the covariance matrix $Q$ does not change for a sufficiently long period. Hence, $\lambda_j, j=0,1,\cdots,2^r-1$ (\ref{eq:lambda}) can be pre-computed and the received signal vectors $\{\mathbf{r}_k\}$ are properly modified to obtain $\{\mathbf{y}_k= Q^{-1}\mathbf{r}_k\}$.

The inner products $\left<\mathbf{h}^j, \mathbf{y}_k\right>, j=0,1,\cdots,2^r-1$ are elements of $H_n \mathbf{y}_k$ and thus can be efficiently evaluated by an FHT of size $n=2^r$. The results are then modified using the pre-computed $\lambda_j$ and $2n$ exponential functions are performed for obtaining $\psi(\pm \mathbf{h}^j), j=0,1,\cdots,2^r-1$.

By defining two $2^r\times 1$ sized compound vectors $\mathbf{a}$ and $\mathbf{b}$ with
\begin{eqnarray}
 \label{eq:ab}
a[j]= \left[\begin{array}{c} \psi(+\mathbf{h}^j) \\
 \psi(- \mathbf{h}^j)
 \end{array}\right], j=0,1,\cdots,2^r-1
\end{eqnarray}
and
\begin{eqnarray}
 \label{eq:ab}
b[i]= \left[\begin{array}{c} \sum_{H[i,j]=\pm 1}\psi(\pm \mathbf{h}^j) \\
 \sum_{H[i,j]=\mp 1}\psi(\pm \mathbf{h}^j)
 \end{array}\right], i=0,1,\cdots,2^r-1,
\end{eqnarray}
it can be again shown \cite{LiPingTH} that
\begin{eqnarray}
 \label{eq:ab}
 \mathbf{b} = Q_n \mathbf{a},
\end{eqnarray}
where $Q_n$ is a $n\times n$ matrix obtained from $H_n$ by the following replacements:
\begin{eqnarray}
 \label{eq:5}
 +1 \rightarrow \left[\begin{array}{cc} 1 & 0 \\
 0 & 1
 \end{array}\right],
  -1 \rightarrow \left[\begin{array}{cc} 0 & 1 \\
 1 & 0
 \end{array}\right].
\end{eqnarray}
The derivation process is the same as that of \cite{LiPingTH}, and thus omitted here.

As shown in \cite{LiPingTH}, the computation of (\ref{eq:ab}) can be efficiently carried out using an $n$-point APP-FHT.

\subsection {Soft-In/Soft-Out Decoding of Convolutional-Hadamard Codes}
For soft-in soft-out (SISO) decoding of convolutional-Hadamard codes, the Bahl, Cocke, Jelinek, and Raviv (BCJR) algorithm can be employed. It is based on the trellis description of convolutional-Hadamard codes as shown in \cite{LiPingTH}. A simple $1/(1+D)$ convolutional code has 2 states and $2^r$ branches per state. Therefore, there are $2^r$ parallel edges connecting $s_k$ (state at the $k$th block-wise time epoch) to $s_{k+1}$ and the branch metric function can be computed as
\setlength{\arraycolsep}{0pt}
\begin{eqnarray}
\label{eq:sd}
   \gamma_k(s_k, s_{k+1}) = \sum_{\mathbf{c}_k\in \mathcal{C}(s_k,s_{k+1})} \psi(\mathbf{c}_k),
\end{eqnarray}
\setlength{\arraycolsep}{5pt}
where $\mathcal{C}(s_k,s_{k+1})$ is the set of possible Hadamard codewords connecting $s_k$ to $s_{k+1}$.

The BCJR algorithm can be characterized by the following forward
and backward recursions:
\begin{eqnarray}
\label{eq:fd}
   \alpha_{k+1}(s_{k+1})= \sum_{s_k} \alpha_k(s_k) \gamma_k(s_k,s_{k+1}),
\end{eqnarray}
\begin{eqnarray}
\label{eq:bd}
   \beta_{k}(s_k)= \sum_{s_{k+1}} \beta_{k+1}(s_{k+1}) \gamma_k(s_k,s_{k+1}).
\end{eqnarray}
Finally, the log-likelihood ratio at the $i$th bit-wise time epoch for the $k$th Hadamard codeword is computed as
\begin{eqnarray}
  \label{eq:13}
  L_k(i) =  \log\frac{\displaystyle\sum_{c_k [i]=+ 1} \psi(\mathbf{c}_k)\alpha_k(s_k) \beta_{k+1}(s_{k+1})}{\displaystyle\sum_{c_k [i]= -1} \psi(\mathbf{c}_k)\alpha_k(s_k) \beta_{k+1}(s_{k+1})}.
\end{eqnarray}

For iterative decoding, it is essential to compute the extrinsic information for systematic information bits. For an AWGN channel, the extrinsic information in the log-likelihood form at the positions of information bits can be computed as
\begin{equation}
\label{eq:le}
   L_k^e(i) = L_k(i) - L_k^a(i) - L_c \cdot r_k(i), i\in \mathcal{I},
\end{equation}
where $L_c = 2\frac{\sqrt{P_s}}{\sigma^2}$.

In the presence of NBI, the last term in (\ref{eq:le}) cannot be directly employed as it was ruined by NBI. Hence, we need to estimate this term, which, however, is not easy in general in the presence of NBI. Here, we propose an approximation by simply omitting this term, giving
\begin{equation}
   L_k^e(i)\approx L_k(i) - L_k^a(i), i\in \mathcal{I},
\end{equation}
for the extrinsic information at the positions of information bits. Indeed, due to very low-rate coding, we have $|\mathcal{I}| \ll L$, and the contribution of $L_c \cdot r_k(i)$ is also very limited. Hence, this approximation is justified. Simulations show that this approximation method does work well.

\subsection{The Overall Decoder}
With the SISO decoder for each component convolutional-Hadamard code, the overall decoder structure is the same as that of Fig. 7 in \cite{LiPingTH}, and thus omitted here.

\section{Extension to Multipath Fading Channels}
In practical wireless communication systems, an AWGN channel model may not be applicable. Instead, multipath fading channels are often considered.

In general, the multipath scenario suffers several difficulties. Firstly, low-rate turbo-Hadamard codes perform well in an AWGN channel, which, however, may perform badly for some multipath channels (or inter-symbol interference (ISI) channels). Secondly, with the introduction of fading, practical wireless communications  essentially see an ensemble of multipath (or ISI) channels, which constitutes a main obstacle for code design. Indeed, design of ``good" codes for multipath fading channels is still a big challenge.

In this section, we do not discuss the code design problem. Instead, we focus on the decoding of low-rate turbo Hadamard codes in the presence of both NBI and multipath fading. Quasi-static fading is considered, where the multipath taps remain fixed during the block of a codeword, but independently varying from one block to another. We assume that the receiver input is preceded by a matched filter, and then followed by a noise whitening filter. Hence, a discrete-time complex baseband representation of the multi-path channel with $P+1$ non-zero taps is considered, where the delay
of tap $l$ is denoted as $\tau_l$, and the coefficient of tap $l$ is $f_l$. The channel output in its complex form can be written as
\begin{eqnarray}
   \label{scl_fd}
    \tilde{r}_k = f_0 c_k + \sum_{l=1}^P f_v c_{k-\tau_l} + \tilde{i}_k + \tilde{w}_k,
\end{eqnarray}
where $\{\tilde{i}_k\}$ is a complex NBI sequence, and $\{\tilde{w}_k\}$ is a complex AWGN sequence.

Suppose the length of Hadamard sequence is large enough so that $\max\{\tau_1,\cdots,\tau_P\}\le n$. By collecting $n$ received samples (\ref{scl_fd}) (similar to (\ref{vec_ch})), the received signal in a vector form of length $n$  can be written as
\begin{eqnarray}
   \label{vec_fd}
    \tilde{\mathbf{r}}_k = f_0 \mathbf{c}_k + \sum_{l=1}^P f_l \bar{\mathbf{c}}_{k-\tau_l}  + \tilde{\mathbf{i}}_k + \tilde{\mathbf{w}}_k,
\end{eqnarray}
where $\bar{\mathbf{c}}_{k-\tau_l}= \mathbf{c}_k^{+\tau_l} + \mathbf{c}_{k-1}^{-(n-\tau_l)}$,  $\mathbf{c}_k^{\pm\tau_l}$ denotes $\mathbf{c}_k$ shifting down (+) or up (-) by $\tau_l$ positions and the vacant positions are filled with zeros.

For simplicity, it is assumed that both $\{f_l\}$ and $\{\tau_l\}$ are perfectly estimated and available at the receiver.   Given $\{f_l\}$ and $\{\tau_l\}$, we denote the ISI part in (\ref{vec_fd}) as
\begin{eqnarray}
\label{fadISI}
    \bar{\mathbf{c}}_{k-1}^k \triangleq \varphi\left(\mathbf{c}_k, \mathbf{c}_{k-1}\right)= f_0 \mathbf{c}_k + \sum_{l=1}^L f_l \tilde{\mathbf{c}}_{k-\tau_l},
\end{eqnarray}
which is a function of both $\mathbf{c}_k$ and $\mathbf{c}_{k-1}$.

In the presence of both NBI and AWGN, one can show that
\begin{eqnarray}
\label{gamm_fad}
     \log p\left(\tilde{\mathbf{r}}_k|\mathbf{c}_k,\mathbf{c}_{k-1}\right) \propto  - \left(\tilde{\mathbf{r}}_k - \bar{\mathbf{c}}_{k-1}^k \right)^\dag \tilde{Q}^{-1}\left(\tilde{\mathbf{r}}_k - \bar{\mathbf{c}}_{k-1}^k \right)
\end{eqnarray}
where $\tilde{Q}= E[ \tilde{\mathbf{i}}_k \tilde{\mathbf{i}}_k^\dag]+E[ \tilde{\mathbf{w}}_k \tilde{\mathbf{w}}_k^\dag]$ with $(\cdot)\dag$ denoting the complex conjugate of the transpose.

The SISO decoding of component convolutional Hadamard codes can now be again implemented using BCJR algorithm, but with expanded super-state trellis. The super state at time epoch $k$ is formed by joining the state pair of the component convolutional Hadamard code at time epoches of $k$ and $k-1$, namely, $s^{sup}_k=(s_{k-1},s_k)$.  With $\max\{\tau_1,\cdots,\tau_L\} \le n$, the multi-path fading essentially introduces the $symbol$ memory of 2 with $symbol$ denoting the vector of Hadamard sequence. Let $S$ denote the number of states for each component convolutional Hadamard code. Then, the number of super-states is $2S$. Therefore, the decoding complexity is almost independent of $P$ thanks to the use of long Hadamard sequences.

With SISO decoding of component convolutional Hadamard codes, iterative decoding of turbo Hadamard codes in the presence of both NBI and multipath fading can be implemented in a similar manner as discussed in Section-IV.

\section{Simulation Results}

In this section, the proposed turbo-Hadamard coding approach is compared with the code-aided DS-SS approach with convolutional coding. We employ the traditional convolutional code instead of the powerful turbo code for the code-aided DS-SS approach, due to the use of short block length for fast simulations. In our simulation, the information bit length of 200 is employed for the convolutional code of constraint length 7, and Viterbi decoding is employed for the NBI suppressed MMSE (soft) outputs $\mathbf{u}^T \mathbf{r}_k, k=1,\cdots,F$ (please refer to (\ref{eq:MMSE})). With this short information bit length, we notice that turbo codes cannot show much advantage over the convolutional code of constraint length 7 in performance.

In the past years, various NBI suppression methods were widely employed such as in the literature \cite{Milstein,AndreaNBI,Yuan,GomaaSparsity,PerezNBI,LamareNBI,SongNBI}.  Although various NBI suppression techniques may result in different performance, we do not compare these techniques with our work, since much of these works focus on the signal processing techniques. By employing the upper bounds for evaluating the ML performance as in Section-II, the traditional DS-SS approach shows its performance weakness compared to a full coding approach for NBI suppression, no matter whatever signal processing technique is employed. This theoretical performance weakness is mainly caused by its spreading (repetition coding) structure, which leaves limited room for performance improvement with advanced signal processing techniques.

For a full turbo-Hadamard code, we use $N$ for the interleaver length, $M$ for the number of component codes, $R$ for the code rate, $r$ for the order of the included Hadamard code, $g(D)$ for the generator polynomial of the employed convolutional code and $\kappa$ for the iteration number. Unless specified otherwise, $\kappa$ is set to 10.

For ease of comparison, it is supposed that the chip rates observed over the channel for both approaches are just the same. For the code-aided DS-SS approach, the spreading sequence is a Hadamard sequence with length $32$ and the rate-1/2 convolutional code with generators $G_1=171$, $G_2=131$ is employed.  For the turbo-Hadamard coding approach, the parameters are set as follows: $M=2$, $r=5$ and $g(D)=\frac{1}{1+D}$. Hence, the Hadamard sequences are of the same length 32 for both approaches.

For fast simulations, $N=200$ is assumed unless specified otherwise.  With various parameters described as above, the rate of the full turbo-Hadamard code is $R_F=\frac{r}{M 2^r}=\frac{5}{64}$.  If information bits are input at a rate of $f_b$ bits/s, the resulting transmission chip rate observed over the channel should be equal to $f_c = f_b /R = f_b M 2^r /r$.  For the rate-1/2 convolutional coded DS-SS system, the transmission chip rate is equal to $f_c = 2f_b 2^r$. For APP decoding of Hadamard codes in the presence of NBI, the covariance matrix $Q$ is required. In simulations, $Q=E\left\{\mathbf{r}_k \mathbf{r}_k^T\right\}$ is pre-computed during the training mode.

With an equal channel bandwidth allocated for both approaches, it is clear that the information transmission rate for the turbo-Hadamard coding approach can be $r=5$ times higher than that of the traditional DS-SS approach with rate-1/2 convolutional coding. In what follows, we show that with such a significantly-higher transmission rate, the turbo-Hadamard coding approach can still perform better than the traditional code-aided DS-SS approach.

\subsection{Tone Interference}

For tone interference, its complex version takes the form of
\begin{eqnarray}
\label{eq:tone}
  \tilde{i}_k = \sum_{l=1}^F \sqrt{P_l} e^{j(2\pi f_l k + \phi_l)},
\end{eqnarray}
where $F$ is the number of tones, $P_l$ and $f_l$ are the power and normalized frequency of
the $l$th sinusoid, and $\{\phi_l\}$ are independent random phases
uniformly distributed over $[0,2\pi)$. For the real channel model employed in (\ref{vec_ch}), it simply takes the real part of $\tilde{i}_k$, namely, $i_k = \Re\left({\tilde{i}_k}\right)$.

\begin{figure}[htbp]
   \centering
   \includegraphics[width=0.5\textwidth]{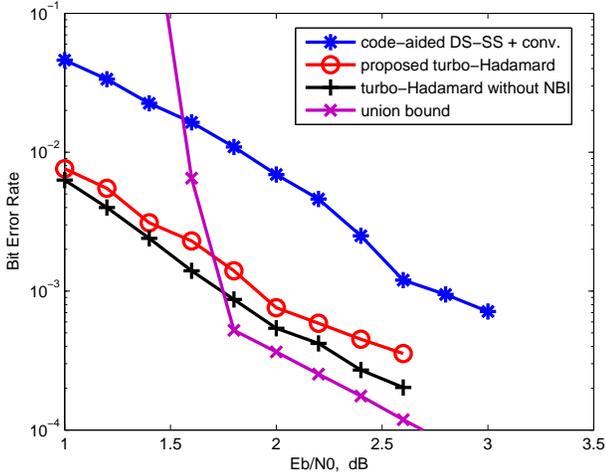}
   \caption{Comparison of the BER performance for the two approaches with multi-tone interference (orthogonal signatures).}
   \label{fig:tone}
\end{figure}
 In \cite{ICCWu}, we have shown the performance of the proposed turbo Hadmard approach under the general multi-tone interference with randomly-generated frequencies. Here, we focus on the multi-tone interference with orthogonal signatures. Simulation conditions are set as follows: $F=3$, the interfering tones have the same power, and the total interference power is kept constant at 25 dB relative to a unity power signal $P_s = 1$. To ensure orthogonal signatures,  $f_l-f_k$ is chosen as the multiple of $1/n=2^{-r}$ for all $l\neq k$, and $\phi_l$ is randomly picked from $[0,2\pi)$. All of conditions remain the same for both approaches.

Simulations have been carried out to show the capability of both approaches against the multi-tone interference with orthogonal signatures. It is shown in Fig. \ref{fig:tone} that the proposed turbo-Hadamard coding approach still outperforms the traditional code-aided DS-SS approach although its transmission rate is about 5 times higher than the code-aided DS-SS approach. In Fig. \ref{fig:tone}, we also plotted the decoding performance for turbo Hadamard codes with perfect knowledge of NBI, as a benchmark. As shown, the proposed turbo-Hadmard coding approach performs very close to the decoding counterpart with perfect knowledge of NBI, and about 0.2 dB in SNR loss is observed.

For multi-tone interference with orthogonal signatures, the union upper bound on the bit error probability similar to (\ref{eq:fer}) is also shown in Fig. 4. By referring to (\ref{eq:fer}), the loss in $E_b/N_0$ is about 0.4 dB due to multi-tone interference. Clearly, the multi-tone interference with orthogonal signatures are much less harmful than the general multitione interference.

\subsection{Autoregressive (AR) Interference}
Suppose now that the NBI signal can be modeled as a $p$-th order AR process, where $p\ll N$, i.e.,
\begin{eqnarray}
\label{eq:ar}
  i_k = -\sum_{j=1}^p a_j i_{k-j} + e_k,
\end{eqnarray}
where $e_k$ is a white Gaussian process with variance $\varepsilon^2$.
\begin{figure}[htbp]
   \centering
   \includegraphics[width=0.5\textwidth]{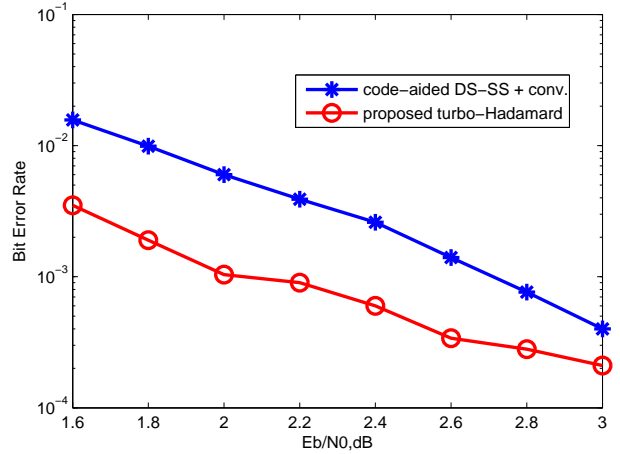}
   \caption{Comparison of the BER performance for the two approaches with AR interference.}
   \label{fig:ar}
\end{figure}

In our simulation, the NBI is generated by a second-order AR model with both poles at
0.99, i.e., $a_1= -1.98$ and  $a_2=0.98$. The total interference power is kept constant at 25 dB relative to a unity power
signal ($P_s = 1$).

It is shown in Fig. \ref{fig:ar} that the proposed turbo-Hadamard coding approach outperforms the traditional code-aided DS-SS approach although its transmission rate is about 5 times higher than the code-aided DS-SS approach.

\subsection{Digital Interference}
\begin{figure}[htbp]
   \centering
   \includegraphics[width=0.5\textwidth]{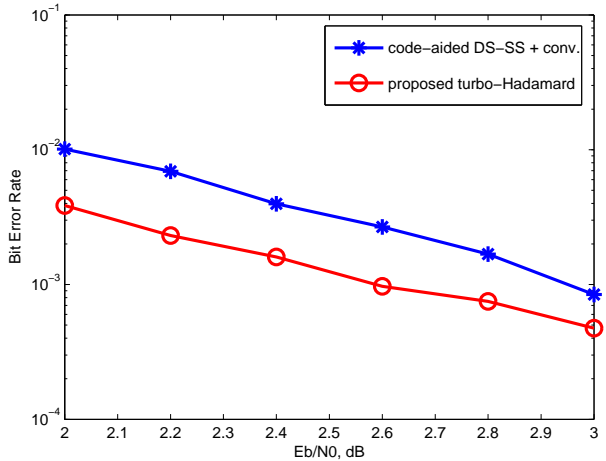}
   \caption{Comparison of the BER performance for the two approaches with digital interference.}
   \label{fig:dig}
\end{figure}

An NBI signal can also be a digital communication signal with a data rate much lower than the spread-spectrum chip rate. The digital NBI model can find its applications when the spread-spectrum communication system works concurrently with some narrow-band digital communication systems.
Here, we assume a fixed relationship between the data rates of the two users, i.e., $q$ bits of the narrow-band user occur for each
bit of the SS user.

In general, the digital NBI signal and the SS signal are asynchronous. Let $t_0$  be the fixed time lag between
the SS bit and the nearest previous start of an NBI bit.

The digital NBI is generated by a base-band digital BPSK signalling with rate $f_{I}=\frac{f_c}{q}$ and a fixed lag $t_0=13 T_c$. The total interference power is kept constant at 20 dB relative to a unity power signal ($P_s = 1$).

Assuming that the digital NBI signal has a waveform of rectangular pulse, the autocorrelation function
of the chip-sampled NBI signal is then
\begin{eqnarray}
\label{eq:dnbi}
  R_i (k)=\left\{\begin{array}{c} \frac{P_i}{L}(1-\frac{|k|}{L}), |k|\le L \\
 0, \quad{}\quad{}\quad{}\quad{}  |k|> L \end{array}\right.
\end{eqnarray}

Although the digital NBI is far from Gaussianly-distributed, the proposed algorithm does work and even outperforms the traditional code-aided DS-SS approach as shown in Fig. \ref{fig:dig}.

\subsection{Impacts of $M$, $N$ and $r$}

\begin{figure}[htbp]
   \centering
   \includegraphics[width=0.5\textwidth]{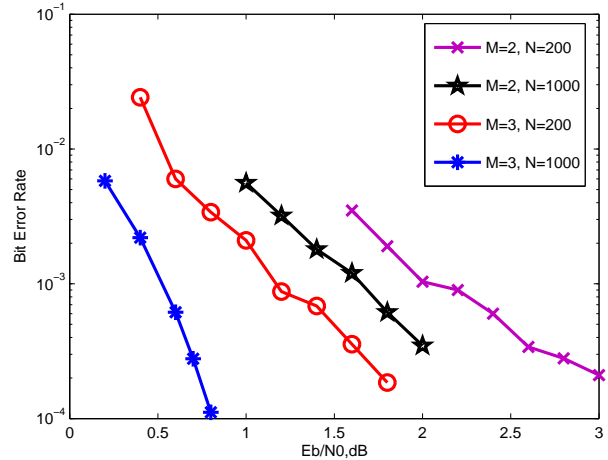}
   \caption{The BER performance of the turbo-Hadamard coding approach for different values of $M$ and $N$ with AR interference.}
   \label{fig:arTHpara}
\end{figure}

The performance of turbo-Hadamard coding approach can be enhanced if either the interleave length $N$ or the number of component codes $M$ increases.
This is well explained in Fig. \ref{fig:arTHpara} for the AR interference with the same setting as before. It should be noted that the increase of $M$ can reduce the spectral efficiency.
As shown, the turbo-Hadamard coding approach performs excellent with $M=3$ and $N=1000$ for strong AR interferences.

\begin{figure}[htbp]
   \centering
   \includegraphics[width=0.5\textwidth]{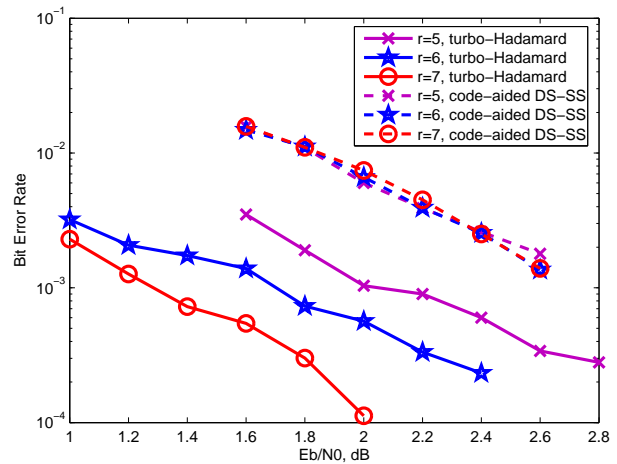}
   \caption{The BER performance of the two approaches for different values of $r$ with AR interference.}
   \label{fig:arTHr}
\end{figure}

With the increase of $r$, the length of spreading sequence is exponentially expanded ($n=2^r$), resulting into improved processing gain. However, the BER performance of the code-aided DS-SS approach with convolutional coding cannot be improved as shown in Fig. \ref{fig:arTHr}. The reason is that the capability of NBI suppression can be achieved with relatively short spreading sequence and its BER performance is largely determined by the employed convolutional coding scheme, which remains almost fixed with the increase of $r$.   With the proposed turbo-Hadamard coding approach, the BER performance can be steadily improved with the increase of $r$, which means that a significant coding gain can be obtained with a lower coding rate.

\subsection{Multipath Fading Channels}
\begin{figure}[htbp]
   \centering
   \includegraphics[width=0.5\textwidth]{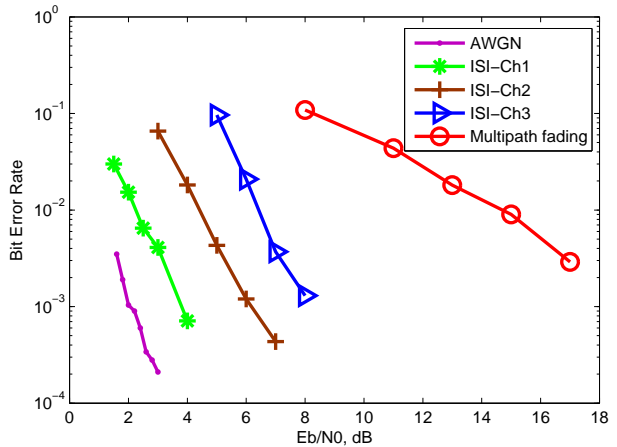}
   \caption{The BER performance of turbo-Hadamard coding approach in both NBI  and multipath (fading) channels, where ISI-Ch1: $[f_0,f_1]=[\sqrt{3/4},\sqrt{1/4}]$; ISI-Ch2:$[f_0,f_1]=[\sqrt{1/2},\sqrt{1/2}]$; ISI-Ch3:$[f_0,f_1]=[\sqrt{1/4},\sqrt{3/4}].$}
   \label{fig:mulpath}
\end{figure}
To gain an essential insight into the performance of the proposed full coding approach under multipath fading channels, a simple two-tap channel model with $\tau_1=n$ is employed. Hence, this ISI channel model can be characterized by $[f_0,f_1]$ if no fading is considered. We consider three ISI channel models, namely, ISI-Ch1: $[f_0,f_1]=[\sqrt{3/4},\sqrt{1/4}]$, ISI-Ch2:$[f_0,f_1]=[\sqrt{1/2},\sqrt{1/2}]$, and ISI-Ch3:$[f_0,f_1]=[\sqrt{1/4},\sqrt{3/4}]$, respectively.

The complex AR interference is assumed, and its real (or imaginary) part takes the same setting as before. As shown in Fig. \ref{fig:mulpath}, the decoding performance under ISI channels perform worse than that under the AWGN channel, and deteriorates when the second tap $f_1$ becomes more dominant.

With quasi-static fading, we consider that both $f_0$ and $f_1$ are independently generated according to complex Gaussian distribution with zero mean and variance of $\frac{1}{2}$. Also shown in Fig. \ref{fig:mulpath}, there is an obvious error floor for multipath fading channel. The error floor comes from some bad realizations of multipath fading channel, namely,  $\max\{|f_1|,|f_0|\} \ll 1$ or $|f_1| \gg |f_0|$.

\subsection{Complexity Issue}
\begin{table}{\bfseries TABLE I: Decoding Complexity Per Information Bit} \\[1ex]
    \begin{tabular} {|c|c|c|c|}
    \hline
    Approach & Mul./Div. & Exp./Log. & Add./Comp. \\
    \hline \hline
    Proposed turbo-Hadamard  & 340 & 276 & 824  \\
    \hline
    Code-aided DS-SS & 2 & 0 & 192 \\
    \hline
    \end{tabular}
\end{table}
In this subsection, we summarize the computational complexity of the proposed turbo-Hadamard coding approach in the AWGN channel, which is compared with that of the code-aided DS-SS approach with convolutional coding. For ease of comparison, we assume that the length of spreading sequence is of size $2^r$, as adopted in our simulation.

With the introduction of NBI, both approaches require to modify the received signal vectors as $\mathbf{y}_k = Q^{-1} \mathbf{r}_k$, where $Q^{-1}$ can be efficiently updated using the recursive least-squares (RLS) adaptive algorithms. Compared to the code-aided DS-SS approach, the proposed approach requires to further pre-compute $\lambda_j, j=0,\cdots, 2^r-1$ (\ref{eq:lambda}). As the cost in this pre-computation stage is almost the same, we do not consider it for final comparison.

The decoding complexity of the proposed turbo-Hadamard coding approach can be summarized as follows. With iterative decoding, the decoding complexity is determined by the number of iterations $\kappa$ and the operations required for $M$ component APP decoders in each decoding iteration. Each convolutional-Hadmard APP decoder is implemented section-by-section over $S$-state trellis and it was estimated in \cite{LiPingTH} that for each section, about $10\times2^r-2r+4S-12$ additions, $2\times 2^r+4+8S$ multiplications/divisions, $2\times2^r+r$ exponential/logrithm functions are required.

The decoding complexity of the code-aided DS-SS approach with convolutional coding is evident \cite{Poor_codeI}. Whenever the refined signal vector $\mathbf{y}_k$ is computed, the linear MMSE detector can be directly implemented by vector multiplications. With rate-1/2 convolutional coding, it requires 2 multiplications per information bit. The decoding complexity of convolutional code is well known, which is determined by the constraint length (Viterbi algorithm).

The decoding complexities per information bit are summarized in Table. I for both approaches ($r=5,M=2,S=2,\kappa=10$). The proposed turbo-Hadmard coding approach has significantly higher complexity, which is mainly due to the full coding approach and the employment of iterative decoding.

\section{Conclusion and Future Work}
In this paper, we have derived upper bounds on the error performance of coded systems in the presence of both AWGN and multi-tone interference with orthogonal signatures, which can be employed to evaluate the system performance. By using tight bounding techniques, we have provided a partial justification that the optimum coding-spreading favors full coding for the purpose of NBI suppression.  It should be emphasized that a rigourous proof of this justification is difficult due to mathematical intractability for general classes of NBI.

More practically, we have proposed a low-rate full turbo-Hadamard coding approach for NBI suppression. With proper iterative processing, it was shown that the proposed low-rate coding approach performs better than the traditional coded DS-SS approach for general classes of NBI with noticeably higher spectral efficiency. For iterative decoding Hadamard codes in the presence of NBI, we found that the fast FHT-based APP algorithm can still work, which makes this approach attractive in practice. The proposed full coding approach can also be adapted to multipath fading channels, and a low-complexity iterative receiver exists if the length of underlying Hadamard sequences is not smaller than the multipath time.

With a full turbo-Hadamard coding approach, the employment of DS-SS signalling with time-varying spreading sequences can be assumed, which could be potentially employed for signal acquisition. Indeed, there are only $n=2^r$ spreading sequences for an $n\times n$ Hadamard matrix with $n=2^r$, while the space of binary spreading sequences of length $n$ is of size $2^n$. Therefore, the receiver may search $n=2^r$ spreading sequences in parallel for the acquisition of the incoming full turbo-Hadmard coded signal, where FHT could be efficiently employed to make this search process faster. As this topic is rather involved, much work remains open. Indeed, we can expect a prospect application of full coding approach in future military communications if the acquisition of low-rate full coding signal can be efficiently implemented.

Finally, the multiuser scenario deserves investigation in the future.

\section*{Acknowledgment}
The authors would like to thank Prof. Xiang-Gen Xia  and Prof. Wei-Ping Zhu for their helpful comments.

\appendix[Evaluation of the PEP]
\newcounter{saveeqn}%
\stepcounter{saveeqn}\setcounter{equation}{0}%
\renewcommand{\theequation}{A.\arabic{equation}}

In this appendix, we consider a special class of NBI process with $F$ tones $f_l, l=1,2,\cdots,F$, as defined by (\ref{eq:tone}).
For ease of derivation, we further assume orthogonal signatures, namely,
\begin{equation}
    \mathbf{g}^H_l \mathbf{g}_k=n \delta_{l,k}, \forall l, k \in [1,F]
\end{equation}
where $\mathbf{g}_l=[1,e^{j2\pi f_l},e^{j4\pi f_l},\cdots,e^{j2\pi (n-1) f_l}]^T$.

It has been shown in \cite{Poor_codeI} that
\begin{equation}\label{eq:f}
  Q^{-1}= \sigma^{-2} \left(I - \sum_{l=0}^F \frac{P_l}{\sigma^2+n P_l}\mathbf{g}_l \mathbf{g}_l^H\right),
\end{equation}

Then, we have
\begin{eqnarray}
   \lambda(\mathbf{c})&=&  - \frac{1}{2 \sigma^2} \sum_{k=1}^B  \left(1-\sum_{l=0}^F \frac{\gamma_l}{1+n \gamma_l} \left|\mathbf{c}_k^T \mathbf{g}_l \right|^2\right) \nonumber \\
    &=&  - \frac{B}{2 \sigma^2}  \left(1-\sum_{l=0}^F \frac{\gamma_l}{1+n \gamma_l} \cdot \frac{1}{B}\sum_{k=1}^B \left|\mathbf{c}_k^T \mathbf{g}_l \right|^2 \right) \nonumber \\
    &\approx&  - \frac{B}{2 \sigma^2}  \left(1-\sum_{l=0}^F \frac{\gamma_l}{1+n \gamma_l} E\left\{\left|\mathbf{c}_k^T \mathbf{g}_l \right|^2\right\} \right) \nonumber \\
    &=&  - \frac{B}{2 \sigma^2}  \left(1-\sum_{l=0}^F \frac{n \gamma_l}{1+n \gamma_l} \right),
\end{eqnarray}
where $\gamma_l = P_l/\sigma^2$, $E\{\cdot\}$ is taken with respect to $\mathbf{c}_k$ and the last step is due to
\begin{eqnarray}
  E\left\{\left|\mathbf{c}_k^T \mathbf{g}_l \right|^2\right\} &=& \text{tr}\left( E\{ \mathbf{c}_k^T \mathbf{c}_k\} \mathbf{g}_l \mathbf{g}_l^H\right)  \nonumber  \\
  &=&  n.
\end{eqnarray}
In the above derviation, we have used the approximation that $\frac{1}{B}\sum_{k=1}^B \left|\mathbf{c}_k^T \mathbf{g}_l \right|^2 \approx E\left\{\left|\mathbf{c}_k^T \mathbf{g}_l \right|^2\right\}$ for large $B$ and the ergodicity of the sub-vector $\mathbf{c}_k$ over the column space of $H_n$.

Hence,
\begin{equation}
\label{ha}
  \lambda(\mathbf{c}) - \lambda(\hat{\mathbf{c}}) \approx 0,
\end{equation}
and the metric difference can be simplified as
\begin{equation}
    \eta - \hat{\eta} \approx \sum_{k=1}^B \left< \mathbf{c}_k - \hat{\mathbf{c}}_k, \mathbf{y}_k\right>.
\end{equation}

Next, we proceed to derive both the mean and the variance of $\eta - \hat{\eta}$ assuming that the codeword $\mathbf{c}$ is transmitted. It is straightforward to show
\begin{eqnarray}
    \mu_z \triangleq E\left\{ \eta - \hat{\eta} \Big| \mathbf{c} \right\} = \sum_{k=1}^B \left(\mathbf{c}_k - \hat{\mathbf{c}}_k \right)^T Q^{-1} \mathbf{c}_k .
\end{eqnarray}

Without loss of generality, we assume that $\mathbf{c}_k$ differs from $\hat{\mathbf{c}}_k$ at $h_k$ head positions. Hence, we employ a sub-vector decomposition of both $\mathbf{c}_k$ and $\hat{\mathbf{c}}_k$, namely,  $\mathbf{c}_k = [\mathbf{u}^T_{h_k}, \mathbf{v}^T_{n-h_k}]^T$ and $\hat{\mathbf{c}}_k = [-\mathbf{u}^T_{h_k}, \mathbf{v}^T_{n-h_k}]^T$.  Thus, $\mathbf{c}_k - \hat{\mathbf{c}}_k =[2\mathbf{u}^T_{h_k}, \mathbf{0}^T_{n-h_k}]^T$.

It follows that
\begin{eqnarray}
  \left(\mathbf{c}_k - \hat{\mathbf{c}}_k\right)^T Q^{-1} \mathbf{c}_k = 2\sigma^{-2} \left(h_k-\sum_{l=0}^F \frac{\gamma_l}{1+n \gamma_l} \xi_{k,l} \right),
\end{eqnarray}
where $\xi_{k,l}= \left|\mathbf{u}^T_{h_k} \mathbf{g}_{l,h_k}\right|^2 + \mathbf{u}^T_{h_k} \mathbf{g}_{l,h_k} \mathbf{g}^H_{l,n-h_k} \mathbf{v}_{n-h_k}$ and $\mathbf{g}_l=[\mathbf{g}^T_{l,h_k},\mathbf{g}^T_{l,n-h_k}]^T$.

Thus,
\begin{eqnarray}
  \mu_z &=& 2\sigma^{-2} \sum_{k=1}^B \left(h_k-\sum_{l=0}^F \frac{\gamma_l}{1+n \gamma_l} \xi_{k,l} \right) \nonumber \\
        &=& 2\sigma^{-2} \left(h-\sum_{l=0}^F \frac{\gamma_l}{1+n \gamma_l}  \sum_{k=1}^B \xi_{k,l} \right) \nonumber \\
        &\approx& 2\sigma^{-2} \left(h-\sum_{l=0}^F \frac{\gamma_l}{1+n \gamma_l} \cdot \sum_{k=1}^B E\{\xi_{k,l}\} \right), \nonumber \\
\end{eqnarray}
where $h=\sum_{k=1}^B h_k$ denotes the Hamming distance between $\mathbf{c}$ and $\hat{\mathbf{c}}$.

Due to the summation of $\xi_{k,l}$ over blocks ($k$), we can use $E\{\xi_{k,l}\}$ in place of $\xi_{k,l}$, where $E\{\cdot\}$ is the expectation with respect to $\mathbf{c}_k$ for a given $h_k$, leading to
\begin{eqnarray}
 E\{\xi_{k,l}\} &=& E\{\left|\mathbf{u}^T_{h_k} \mathbf{g}_{l,h_k}\right|^2\} \nonumber \\
   &&+ E\{\mathbf{u}^T_{h_k} \mathbf{g}_{l,h_k}\} \cdot E\{\mathbf{g}^H_{l,n-h_k} \mathbf{v}_{n-h_k}\} \nonumber \\
   &=&h_k.
\end{eqnarray}

Hence, we get
\begin{eqnarray}
  \mu_z \approx& 2\sigma^{-2} \left(1-\sum_{l=0}^F \frac{\gamma_l}{1+n \gamma_l} \right) h.
\end{eqnarray}

For computing the variance, we note that
\setlength{\arraycolsep}{0.0em}
\begin{eqnarray*}
    E \{(\hat{\eta} &-& \eta)^2 \Big| \mathbf{c} \} = \sum_{k=1}^B E\left\{\left|\left< \hat{\mathbf{c}}_k-\mathbf{c}_k, \mathbf{y}_k\right> \right|^2 \Big| \mathbf{c}_k\right\}  \nonumber \\
    &=& \sum_{k=1}^B (\hat{\mathbf{c}}_k-\mathbf{c}_k)^T Q^{-1} \left[\mathbf{c}_k \mathbf{c}^T_k + Q\right] Q^{-1} (\hat{\mathbf{c}}_k-\mathbf{c}_k) \nonumber \\
    &=& E^2\left\{\hat{\eta} - \eta\Big| \mathbf{c} \right\} +  \sum_{k=1}^B(\hat{\mathbf{c}}_k-\mathbf{c}_k)^T Q^{-1} (\hat{\mathbf{c}}_k-\mathbf{c}_k).
\end{eqnarray*}
\setlength{\arraycolsep}{5pt}

Then, we get
\begin{eqnarray}
    \sigma^2_z &\triangleq &D\left\{\hat{\eta} - \eta\Big| \mathbf{c} \right\} = E\left\{(\hat{\eta} - \eta)^2\Big| \mathbf{c} \right\} - E^2\left\{\hat{\eta} - \eta\Big| \mathbf{c} \right\} \nonumber \\
       &=&\sum_{k=1}^B(\hat{\mathbf{c}}_k-\mathbf{c}_k)^T Q^{-1} (\hat{\mathbf{c}}_k-\mathbf{c}_k) \nonumber \\
        &\approx & 4\sigma^{-2}\left(1-\sum_{l=0}^F \frac{\gamma_l}{1+n \gamma_l}\right) h.
\end{eqnarray}



\begin{biography}[{\includegraphics[width=1in,height=1.25in,clip,keepaspectratio]{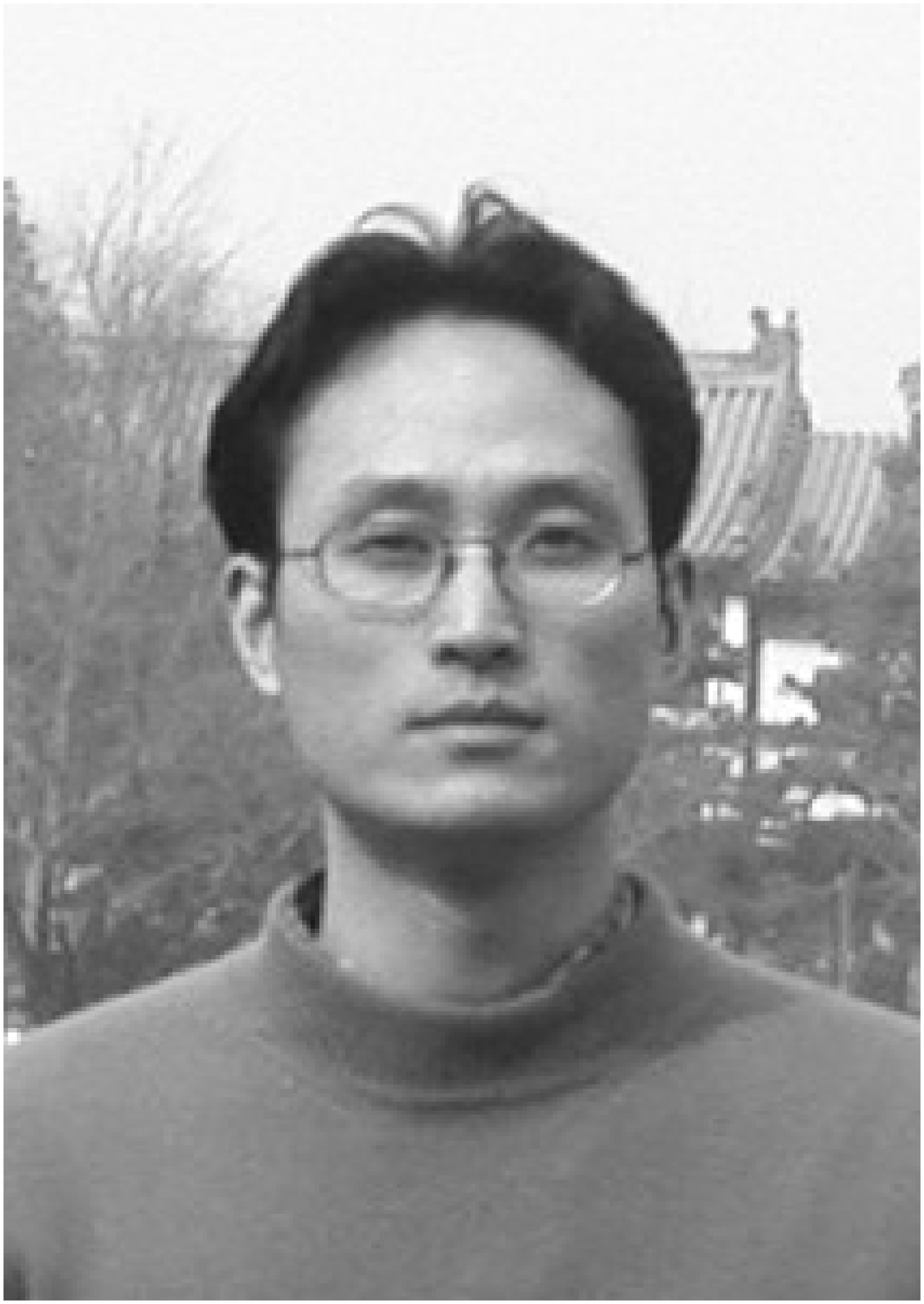}}]{Xiaofu Wu}
was born in Jan. 1975. He received the B.S. and M.S.
degrees in electrical engineering from the Nanjing Institute of
Communications Engineering, Nanjing, China, in 1996 and 1999,
respectively, and the Ph.D. degree in electrical engineering from
the Peking University, Beijing, China, in 2005. From 2005 to 2007, he was with the Southeast University as a Post-Doctoral researcher
 at the National Mobile Communication Research Laboratory.
Since 2012, he has been with the Nanjing University of Posts \& Telecommunications, where he is currently a Professor.

His research interests are in coding and information theory, information-theoretic security and mobile computing. He has contributed over 25 IEEE/IEE jounal papers.
\end{biography}

\begin{biography}[{\includegraphics[width=1in,height=1.25in,clip,keepaspectratio]{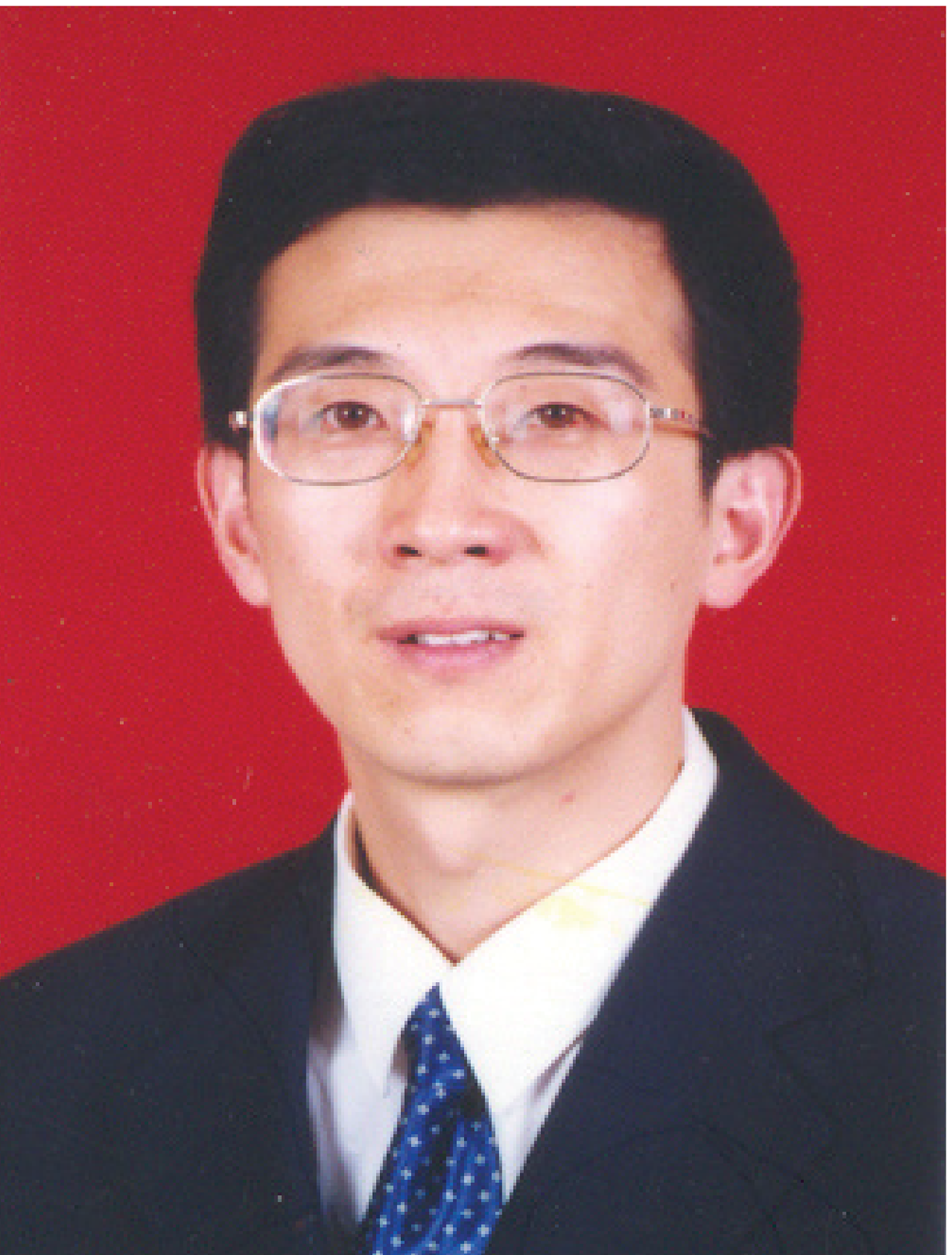}}]{Zhen Yang}
 received his B.Eng. and M.Eng. degrees from Nanjing University of Posts and Telecommunications in 1983 and 1988, respectively, and his Ph.D. degree from Shanghai Jiao Tong University in 1999, all in electrical engineering. He was initially employed as a Lecturer in 1983 by Nanjing University of Posts and Telecommunications, where he was promoted to Associate Professor in 1995 and then a Full Professor in 2000. He  was a visiting scholar at Bremen University, Germany during 1992-1993, and an exchange scholar at Maryland University, USA in 2003. His research interests include various aspects of signal processing and communication, such as communication systems and networks, cognitive radio, spectrum sensing, speech and audio processing, compressive sensing and wireless communication. He has published more than 200 papers in academic journals and conferences.

Prof. Yang serves as Vice Chairman and Fellow of Chinese Institute of Communications, Chairman of Jiangsu Institute of Communications,   Vice Director of Editorial Board of The Journal on Communications and China Communications. He is also a Member of Editorial Board for several other journals such as Chinese Journal of Electronics, Data Collection and Processing et al, the Chair of APCC (Asian Pacific Communication Conference) Steering Committee during 2013~2014.
\end{biography}

\end{document}